\documentclass[12pt]{article}

\usepackage{graphics}
\usepackage{color}
\usepackage{amssymb}
\usepackage{latexsym}

\font\manual=manfnt \def\dbend{\lower3.5pt\hbox{\manual\char127}}

\def\ie{{\it i.e.}}

\def\cf{{\it c.f.}}

\def\etc{{\it etc}}

\def\sst{\scriptscriptstyle}

\def\frac#1#2{{#1\over#2}}
\def\coeff#1#2{{\textstyle{#1\over #2}}}
\def\half{\frac12}
\def\hf{{\textstyle\half}}

\def\IR{{\mathbb R}}
\def\IC{{\mathbb C}}

\def\IZ{{\mathbb Z}}

\catcode`\@=11
\def\slash#1{\mathord{\mathpalette\c@ncel{#1}}}
\def\underrel#1\over#2{\mathrel{\mathop{\kern\z@#1}\limits_{#2}}}

\catcode`\@=12
\def\ket#1{|#1\rangle}

\def\vev#1{\langle#1\rangle}

\def\Tr{{\rm Tr}}

\def\sinh{{\rm sinh}} 	
\def\cosh{{\rm cosh}}

\def\exp{{\rm exp}}

\def\HH{{\cal H}} 
 
\def\JJ{{\cal J}} 
 
\def\LL{{\cal L}} 
 
\def\NN{{\cal N}} 
\def\OO{{\cal O}}

\def\TT{{\cal T}}

\def\unlockat{\catcode`\@=11}
\def\lockat{\catcode`\@=12}
\unlockat
\def\newsec#1{\global\advance\secno by1\message{(\the\secno. #1)}
\global\subsecno=0\global\subsubsecno=0\eqnres@t\noindent
{\bf\the\secno. #1}
\writetoca{{\secsym} {#1}}\par\nobreak\medskip\nobreak}
\global\newcount\subsecno \global\subsecno=0
\def\subsec#1{\global\advance\subsecno
by1\message{(\secsym\the\subsecno. #1)}
\ifnum\lastpenalty>9000\else\bigbreak\fi\global\subsubsecno=0
\noindent{\it\secsym\the\subsecno. #1}
\writetoca{\string\quad {\secsym\the\subsecno.} {#1}}
\par\nobreak\medskip\nobreak}
\global\newcount\subsubsecno \global\subsubsecno=0
\def\subsubsec#1{\global\advance\subsubsecno by1
\message{(\secsym\the\subsecno.\the\subsubsecno. #1)}
\ifnum\lastpenalty>9000\else\bigbreak\fi
\noindent\quad{\secsym\the\subsecno.\the\subsubsecno.}{#1}
\writetoca{\string\qquad{\secsym\the\subsecno.\the\subsubsecno.}{#1}}
\par\nobreak\medskip\nobreak}
\def\subsubseclab#1{\DefWarn#1\xdef
#1{\noexpand\hyperref{}{subsubsection}%
{\secsym\the\subsecno.\the\subsubsecno}%
{\secsym\the\subsecno.\the\subsubsecno}}%
\writedef{#1\leftbracket#1}\wrlabeL{#1=#1}}
\lockat
\input{epsf}

\newcommand{\txfc}[2]{{\textstyle{\frac{#1}{#2}}}}

\newcommand{\nnmb}{\nonumber}

\definecolor{wtmgreen}{rgb}{0,0.4,0}

\newcommand{\be}{\begin{equation}}
\newcommand{\ee}{\end{equation}}

\newcommand{\bbb}{\begin{eqnarray}}
\newcommand{\eee}{\end{eqnarray}}
\newcommand{\pref}[1]{(\ref{#1})}

\begin{document}

%

\def\globalAdS{1}
\def\orbAdS{2}
\def\altorbAdS{3}
\def\znorb{4}
\def\alttrueBTZregion{5}
\def\BTZslices{6}
\def\crunch{7}

\def\ads{AdS}
\def\btz{{\sst BTZ}}
\def\mbar{{\bar m}}
\def\N{{\bf N}}
\def\str{{\sst\rm str}}
\def\gstr{g_s}
\def\lstr{\ell_{s}}
\def\lpl{\ell_{p}}
\def\rads{\ell}
\def\rp{r_+}
\def\rminus{r_-}
\def\rhat{{\hat r}}
\def\that{{\hat t}}
\def\xhat{{\hat x}}
\def\phihat{{\hat \phi}}

\def\Ft{\tilde F}
\def\ttil{\tilde t}
\def\xt{\tilde x}
\def\ut{\tilde u}
\def\vt{\tilde v}
\def\wt{\tilde w}
\def\zt{\tilde z}
\def\At{\tilde A}
\def\Ht{\tilde H}
\def\Kt{\tilde K}
\def\Rt{\tilde R}

\def\tLL{{\tilde\LL}}

\def\osc{{\rm osc}}


\begin{titlepage}
\rightline{EFI-02-85}

\rightline{hep-th/0206175}

\vskip 3cm
\centerline{\Large{\bf Exciting AdS Orbifolds}}

\vskip 2cm
\centerline{
Emil J. Martinec\footnote{\texttt{e-martinec@uchicago.edu}}~~ and ~~
Will McElgin\footnote{\texttt{wmcelgin@theory.uchicago.edu}}}
\vskip 12pt
\centerline{\sl Enrico Fermi Inst. and Dept. of Physics}
\centerline{\sl University of Chicago}
\centerline{\sl 5640 S. Ellis Ave., Chicago, IL 60637, USA}

\vskip 2cm

\begin{abstract}

The supersymmetric $(AdS_3\times S^3)/\IZ_N$ orbifold
constructed by the authors in hep-th/0106171 is shown to describe
AdS fragmentation, where fivebranes are emerging
from the F1-NS5 background.  The twisted sector moduli
of the orbifold are the collective coordinates of
groups of $n_5/N$ fivebranes.  We discuss the relation
between the descriptions of this background as a perturbative
string orbifold and as a BPS state in the dual spacetime CFT.
Finally, we attempt to apply the lessons learned to
the description of BTZ black holes as $AdS_3$ orbifolds
and to related big crunch/big bang cosmological scenarios.

\end{abstract}

\end{titlepage}

\newpage

\setcounter{page}{1}


\section{Introduction and Summary}

Much interesting structure of string theory
has been revealed through the investigation of
its asymptotically $AdS$ backgrounds --
the precise counting of black hole microstates%
~\cite{Strominger:1996sh};
the matching of low-energy black hole emission/absorption spectra 
between string theory and supergravity%
~\cite{Callan:1996dv,Das:1996wn}; 
the UV/IR connection%
~\cite{Susskind:1998dq};
and spacetime topology changing transitions%
~\cite{Witten:1998zw}-\cite{Dijkgraaf:2000fq},
to name but a few.
The overarching idea encompassing all of these results is 
the duality between quantum gravity and 
a holographically dual conformal quantum field theory%
~\cite{Maldacena:1998re}.

The $AdS_3$ example has figured prominently in these
discussions; it arises as the near-horizon geometry
of $n_1$ strings and $n_5$ fivebranes.
In addition to the above list of results,
it has also proved a fruitful ground for exploring
additional issues, such as the black hole information paradox
(see~\cite{Horowitz:1997}-\cite{Maldacena:2001kr}
for a few examples),
and the role of singular conformal field theories%
~\cite{Seiberg:1999xz}.

One distinctive feature of $AdS_3$ is that the supergravity
side of the correspondence admits a perturbative string description
at points in moduli space with vanishing RR fields%
~\cite{Giveon:1998ns}, giving one access to more than just
the low-energy supergravity spectrum and dynamics
on the `bulk' side of the correspondence.
Another is that the current algebraic nature of the
conformal symmetry provides more powerful tools 
in the analysis of the `boundary' theory -- 
in this case a sigma model on the moduli space of instantons%
~\cite{Dijkgraaf:1997nb,Witten:1997yu,Dijkgraaf:1998gf}.

Given the degree of control one has over both
sides of the correspondence for $AdS_3/CFT_2$,
one may ask whether it is possible to study black hole
physics while having simultaneous control over both
a bulk and boundary description of the dynamics,
for instance to see the `long string modes'
describing black hole states~\cite{Maldacena:1996ds}
in a description that also has control over local physics
in the bulk, such as that of~\cite{Giveon:1998ns}.
It was with this goal in mind that the authors 
began a study of $AdS_3\times S^3$ orbifolds
in~\cite{Martinec:2001cf}, in part because of the
naive interpretation of the $AdS_3$ or BTZ black hole%
~\cite{Banados:1992wn,Banados:1993gq} as an orbifold
of global $AdS_3$ spacetime.  

The orbifold procedure
is a useful way of generating new backgrounds from
solvable examples.  Where the orbifold operation takes
the theory seems to be strongly example-dependent.
For example, a chiral $\IZ_N$ orbifold 
of the $S^3$ has the interpretation of adding KK monopole
charge to the background~\cite{Kutasov:1998zh}.
The effect on the dual CFT of this orbifold
operation is not known (see~\cite{Larsen:1999dh} for a discussion).

In~\cite{Martinec:2001cf}, it was shown that the bulk orbifold 
$(AdS_3\times S^3)/\IZ_N$ 
(by opposite $\IZ_N$ rotations in both $S^3$ and in
the spatial directions of $AdS_3$)
yields a BPS state in the {\it same} spacetime CFT 
that describes the original global $AdS_3$ spacetime
as its $SL(2)$ invariant vacuum state. 
Reasons for considering this orbifold include 
(i) the technology for time-independent
orbifolds is well-understood; and 
(ii) naively the geometry approaches the extremal 
BTZ black hole threshold from below as $N\to\infty$,
so one might hope to have a situation where there is
some degree of control over both geometry and 
black hole degrees of freedom.

In this work, we examine the latter construction in
more detail, in particular we find 
in section~\ref{uduality} precisely which 
among a large degeneracy of BPS states 
is being described by the bulk orbifold.
It will turn out that the configuration being described
is that of $N$ groups of $n_5/N$ fivebranes near
the point where they emerge onto the Coulomb branch
of their moduli space%
\footnote{The Coulomb branch for fivebranes 
is lifted at generic points
of the $O(5,4)/O(5)\times O(4)$ moduli space of supergravity
in the F1-NS5 background, but it is present at
the locus of vanishing RR fields described by the
perturbative string formalism of~\cite{Giveon:1998ns}.
This locus of moduli space characterizes the singular CFT's of%
~\cite{Maldacena:1998uz,Seiberg:1999xz,Larsen:1999dh,Giveon:1998ns}.}.
We further give an interpretation of
the various excitations of the orbifold
in section~\ref{excitations} -- 
especially the twisted sectors, where the twisted sector
moduli are shown to describe the relative separations
of the fivebranes.  

It also turns out that there is a remarkable parallel
between the description of the rotational orbifold 
in the bulk perturbative string theory
and the description of the twist vertex operators in the dual
boundary CFT that create the corresponding BPS states
from the $SL(2)$ invariant vacuum.  Namely,
the twist operator correlation functions
have the same geometrical description in terms of 
the vacuum-to-vacuum amplitude on a
branched cover of the Riemann sphere 
(the Euclidean $AdS_3$ boundary), as the bulk orbifold
does in terms of its branched cover onto
the global $AdS_3$ spacetime!

After a brief comment on related fivebrane backgrounds
in section~\ref{fivebranes}, we turn in section~\ref{btzsection}
to a discussion of the $AdS_3$ black hole geometry as an orbifold%
~\cite{Banados:1992wn,Banados:1993gq}.
After reviewing the geometry of the identification, 
we discuss the nature of the vacuum and in particular
its behavior near the black hole singularities.

As a classical geometry, the eternal $AdS_3$ (BTZ) 
black hole is a quotient of the $AdS_3$ geometry
by an element of the $SL(2,\IR)\times SL(2,\IR)$ isometry group 
that roughly speaking acts as a boost rather than a rotation.
Thus, a freely falling observer 
experiences a geometry that near the singularity
looks like a sort of Milne universe.
Such geometries have been periodically identified
as interesting subjects for investigation,
both in the context of perturbative string theory
and in quantum cosmology%
~\cite{%
Horowitz:1991ap}%
-\cite{%
Elitzur:2002rt%
}.
The group action identifies the boundary of $AdS_3$ 
in such a way that the intersection of the black/white hole
singularities with the boundary are again locally
Milne singularities, 
now in the 1+1 dimensional CFT on the Lorentzian cylinder
which is the conformal boundary of $AdS_3$.
If one considers $SL(2,\IR)$ rather than its covering space,
there are four such singularities (two if one further
identifies to the Poincar\'e patch $PSL(2,\IR)$).

The advantage of an $AdS_3$ based example is again that
one may be able to bring to bear both a perturbative
bulk (string) description of the BTZ orbifold~\cite{%
Horowitz:1991ap,%
Satoh:1997xf,%
Satoh:1998xe,%
Cornalba:2002fi,%
Nekrasov:2002kf,%
Simon:2002ma,%
Liu:2002ft%
}
as well as a two-dimensional boundary CFT description.
As discussed above, in matching the two dual descriptions, 
one must overcome the obstacle 
that it is not always known how the orbifold
procedure affects the boundary theory.
This is especially true when we abandon orbifold actions
that preserve sufficient supersymmetry to guarantee
some amount of nonrenormalization of the background,
since we lose control over what properties
of tree-level string theory are maintained in
the full nonperturbative quantum state;
there is no guarantee that string perturbation
theory is an accurate guide to the exact geometry.
Quantum corrections could completely
invalidate the picture of the geometry as an orbifold,
\ie\ as something that can be covered onto the global
$AdS_3$ geometry.
Nevertheless, we proceed under the assumption that states
in the boundary theory which respect the appropriate discrete
symmetries will possess the same basic characteristics as the 
Kruskal vacuum of the BTZ geometry, and attempt to deduce
some of their properties.%
\footnote{The same assumption is implicit in the
work of~\cite{Horowitz:1998xk}, where even more exotic
identifications of $AdS_3$ are considered in the context
of cosmology.  The discrete subgroups of $SL(2,\IR)$
considered there are allowed to act discontinuously
on the conformal boundary -- the parameter space
of the boundary CFT is a noncommutative geometry 
(G. Horowitz and M. Douglas, unpublished).}
If this assumption proves incorrect, then one cannot
characterize the BTZ black hole as an $AdS_3$ orbifold.

A persistent question in the AdS/CFT correspondence 
has been how to describe propagation into the black hole
singularity within the framework of the dual CFT;
a related source of confusion is what it means to propagate
through the horizon of the extremal brane geometry
according to the global time rather than the Poincar\'e
time naturally related to the brane Hamiltonian
(see for example~\cite{Banks:1998dd}).
A fundamental issue is how to treat the intersection
of the horizon (and/or the singularity) with the conformal boundary.
In fact, it is generally true that the outer horizon, inner horizon,
and singularity of any BTZ black hole
all coincide at the boundary of the $AdS$ covering space.
The transformation between the global and
Poincar\'e coordinates, or between global and BTZ
Kruskal coordinates, has singularities at these points;
blithely adding the singular locus to the 
conformal compactification
of the Poincar\'e patch or the Kruskal boundary
is not always justified.
In the present context,
the question becomes whether the states in the boundary
theory that respect the (extremal or
nonextremal) BTZ identification have a sensible description
in the Hilbert space of eigenstates of the global time,
so that one can make sense of propagating them
through the Milne singularities.
A reasonable prescription for dealing with such singularities
in 1+1 dimensional CFT would go a long way toward
justifying recent speculations about cosmological
singularities by providing them with an
interpretation in a fully quantum mechanical
realization of gravity, via the interpretation
of the CFT as the boundary dual to asymptotically 
AdS quantum gravity.

We will find that the stress tensor of the required
CFT state diverges rather badly along the light cones emanating
from the singular locus on the boundary, presenting a
difficulty for the orbifold characterization of
the black hole, and associated cosmological models.  
Such a divergence might be regarded as an instance of the 
generic divergence of stress tensors on Cauchy horizons
(\cf~\cite{Flanagan:1997er} and references therein).

We conclude this introduction
with a comment on the apparent diversity
of presentations of the BTZ black hole.
There are by now many descriptions of CFT states
purportedly dual to the black hole -- the thermal
state in the Hilbert space $\HH$ of the CFT; 
the entangled state in $\HH^{\otimes 2}$ suggested
by the Kruskal geometry; states formed by collapse
of energetic states in supergravity; and the
hyperbolic orbifold.  Are they all the same black hole
geometry?  Our view on this issue is that they are not.
They are distinguished by the history along the entire
spacetime boundary which they possess.
For instance, a generic state in the thermal ensemble,
or one formed by the collapse of energetic probes dropped
from the $AdS$ boundary, employ a single CFT Hilbert space
$\HH$ and describe a spacetime with one asymptotic boundary;
while the correlated state in $\HH^{\otimes 2}$
of~\cite{Maldacena:2001kr}
describes a spacetime with two disjoint asymptopia.  
Any set of observations made on a single asymptotic boundary in 
any of these examples will find what semiclassically
looks like a black hole.
There is a certain artificiality to the AdS/CFT
correspondence when viewed cosmologically, in
that there is an external agent who sets up
the `universe' under consideration.  
The generic high energy eigenstate state in $\HH$
will be approximately thermal in its properties, 
and does not arise from a collapse process, 
but still looks like a black hole.
The universe one `creates' may have one asymptotic boundary, 
as in the case of the thermal CFT state; or it may have two, 
as in the case of the entangled state of the Kruskal extension 
(it may also have many, or possibly none~\cite{Horowitz:1998xk}!).
It may not be possible to detect the full geometry
on the basis of experiments available to any
single observer, but that does not invalidate the
cosmology until and unless one can develop criteria
to eliminate these more exotic examples with 
multiple holographic screens.  This observation may also
have consequences for de Sitter cosmology, where the
supposition of a holographic dual containing only a
horizon area's worth of states assumes that there
is no part of the world holographically inaccessible
to a given observer.
Finally, the ability to cover cosmologies with
closed spatial geometry onto global AdS spacetime
may indicate the presence of non-obvious and/or non-local
observables (\cf~\cite{Elitzur:2002rt}),
whose apparent lack has long been a stumbling block
in understanding closed quantum cosmologies.
The answer to these questions bear upon whether
intial conditions with cosmological past horizons
are admissible, and if so, what data describes the correlations
between various causally disconnected domains.

{\sl Note added.} While this manuscript was in its final stages
of preparation, we learned of related work~\cite{Lunin:2002fw}
on the conical geometries of sections 2 and 3.


\section{\label{uduality}Rotational orbifolds and U-duality}

To begin, we consider the rotational orbifold
$(AdS_3\times S^3)/\IZ_N$.
Our goal is to establish the relation of the orbifold geometry 
with a specific BPS state of the dual spacetime CFT.
By U-duality, the BPS charges of the $n_1$ fundamental strings
and $n_5$ fivebranes wrapped on $T^4\times S^1$ can be mapped
to momentum and winding charges of a fundamental string
on the $S^1$; the BPS configurations of such a string
are well-known to correspond to the different ways that
one can satisfy the level-matching constraints by
exciting left-moving oscillators while keeping the
right-movers in their ground state.
Lunin and Mathur~\cite{Lunin:2001fv,Lunin:2001jy}
have worked out the relation via U-duality of 
the metric sourced by such a fundamental
string carrying winding and momentum on a circle 
(call it the $\xt_5$ direction) to the corresponding
geometry of the D1-D5 or equivalently the F1-NS5 system;
let us recall their analysis.

The chain of dualities in question is
\be
 \mbox{\ \rm F1-NS5\ }~{\buildrel{T_5}\over{\longrightarrow}}~
 \mbox{\ \rm P-NS5\ }~{\buildrel{{\rm 9/11~flip}}\over{\longrightarrow}}~
 \mbox{\ \rm P-D4\ }~{\buildrel{T_{678}}\over{\longrightarrow}}~
 \mbox{\ \rm P-D1\ }~{\buildrel{S}\over{\longrightarrow}}~
 \mbox{\ \rm P-F1\ }
\label{dualchain}
\ee
(we assume that the NS5 brane is additionally wrapped
on a $T^4$ along the 6789 directions).
The BPS charges carried by the U-dual fundamental string
are thus $n_w=n_5$ and $n_p=n_1$.  
As discussed in~\cite{Martinec:1999sa}, 
the P-F1 duality frame is the appropriate 
low-energy description for the very-near 
horizon physics (\ie\ ultra-low energy perturbations),
and so it is appropriate to consider this U-duality
frame for the purpose of understanding the structure of
the vacuum and its low-lying excitations.

Level matching on the string requires
a left-moving oscillator excitation of level $n_1n_5$.
The general class of such states is
(referring dimensions to the string scale $\lstr$,
which will be suppressed)
\bbb
  \xt_i(\tau+\sigma)&=&\sum_m \frac{a_m}{\sqrt m}\;
  	e^{im(\tau+\sigma)}+{\it c.c}\nonumber\\
  \xt_5&=&\frac{n_1}{\Rt}\tau+n_5\Rt\sigma
\label{strconfig} \\
  \ttil&=&E\tau \nonumber
\eee
where classically the amplitude of the oscillation
is the square root of the occupation number $|a_m|\sim \sqrt{N_m}$.
There are of order $exp[2\pi\sqrt{2n_1n_5}]$ possible choices
of oscillator excitation.
The oscillator excitations may be classified according to
representations of the symmetric group, and thus matched
to the BPS states of the F1-NS5 or D1-D5 system,
which are the orbifold cohomology classes of 
$(T^4)^{n_1n_5}/S_{n_1n_5}$~\cite{Vafa:1996zh,Maldacena:1998bw}.


The choice of oscillator excitation
that is related to the $(AdS_3\times S^3)/\IZ_N$
orbifold of \cite{Martinec:2001cf} is
\be
  (\alpha_{-N})^{n_1n_5/N}\ ,
\label{oscstate}
\ee
where the polarization of the oscillators will be determined shortly.
In particular, $N=1$ describes 
the $SL(2)$ invariant vacuum, or global $AdS$.
We wish to see that the metric generated by this oscillator
state U-dualizes to the orbifold metric%
\footnote{Recall that in the F1-NS5 duality frame, 
the $AdS$ scale is $\rads=\sqrt{n_5}\lstr$.  In what follows,
$AdS_3\times S^3$ coordinates will be made dimensionless
by referring them to this scale.}%
~\cite{Martinec:2001cf}
\bbb
ds^{2}/\rads^2 & = & -(r^{2}+N^{-2})\,dt^{2}+(r^{2}+N^{-2})^{-1}\,dr^{2}+
r^{2}d\phi^{2} \nnmb \\[0.2cm]
& & +d\theta^{2}+\cos^{2}\theta\,d\chi^{2}+
\sin^{2}\theta\,(d\psi-N^{-1}d\phi)^{2} \quad .
\label{orbmet}
\eee
This latter metric describes a state in the spacetime CFT
with the quantum numbers (in the NS sector)
\be
  \LL_0+\coeff{c}{24}=\coeff{c}{12}(1-N^{-1})=\TT_0^3
\label{quantnos}
\ee
where $\TT_0^3$ is the left-handed $S^3$ angular momentum,
and $c=6n_1n_5$.  The right-moving quantum numbers
are identical.

\begin{figure}[ht]
\begin{center}
\[
\mbox{\begin{picture}(200,200)(0,0)
\includegraphics{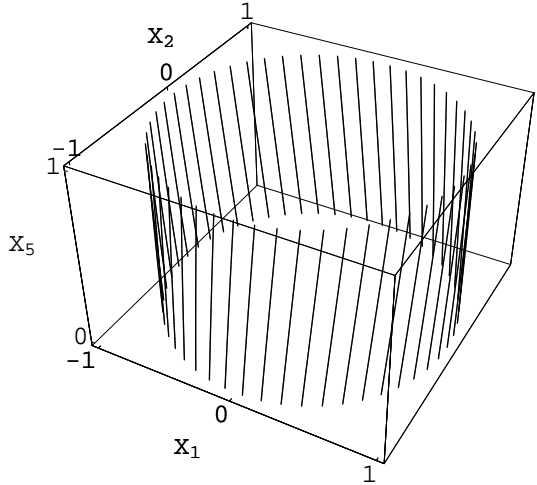}
\end{picture}} 
\]
\caption{\it
The spatial configuration of the U-dual string source for
$n_5=50$, $N=1$, corresponding to global $AdS_3\times S^3$.  
The $\xt_5$ direction is periodically identified
to make an $n_5$ times wound string.  Smearing
the source along $\xt_5$ generates a ring source
in the $\xt_1$-$\xt_2$ plane.
}
\end{center}
\end{figure}

\begin{figure}[ht]
\begin{center}
\[
\mbox{\begin{picture}(200,200)(0,0)
\includegraphics{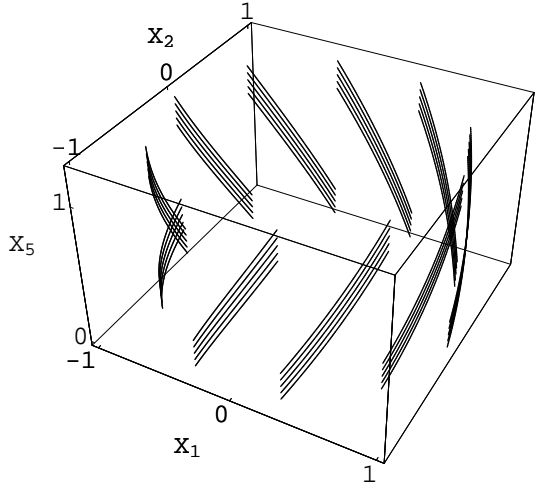}
\end{picture}} 
\]
\caption{\it
The spatial configuration of a U-dual string source for
$n_5=50$, $N=5$, related to $(AdS_3\times S^3)/\IZ_5$.  
The strands are separated slightly for visualization purposes.
The actual source for the orbifold puts the strands at
finite separation.
}
\end{center}
\end{figure}

The $SO(4)=SU(2)\times SU(2)$
angular momentum along the transverse directions 1234
in the system is
$\TT_0^3=\hf(M_{12}+M_{34})$, $\bar\TT_0^3=\hf(M_{12}-M_{34})$.
When the F1-NS5 system is mapped to the P-F1 duality frame,
angular momentum $(-\hf,-\hf)$ for each oscillator
describes a string rotating in the 1-2 plane.
The total angular momentum of
the state~\pref{oscstate} is thus $\TT_0^3=\bar\TT_0^3=n_1n_5/2N$
in R sector conventions, or $\hf n_1n_5(1-1/N)$ in NS sector
conventions (thus matching~\pref{quantnos}).  
Eliminating $\tau$ and $\sigma$ from~\pref{strconfig}
(and solving the Virasoro constraints)
gives the location in the target space of the string source:
\bbb
  \xt_1&\equiv&\Ft_1(\ttil+\xt_5)=
	a\cos\Bigl[\frac{N}{\Rt n_5}(\ttil+\xt_5)\Bigr]\nonumber\\
  \xt_2&\equiv&\Ft_2(\ttil+\xt_5)=
	a\sin\Bigl[\frac{N}{\Rt n_5}(\ttil+\xt_5)\Bigr]
\label{sourceconfig}
\eee
with the amplitude $a=\sqrt{n_1n_5}/N$.
Plots of the spacetime configuration of such a source 
in the P-F1 duality frame are shown for global $AdS_3\times S^3$,
$N=1$, in figure~{\globalAdS}; 
and a naive configuration for $N>1$ is shown in in figure~{\orbAdS}.
Note that the locus of the string in spacetime is traversed 
$N$ times as $\sigma$ goes from zero to $2\pi$.

The geometry sourced by this string configuration is
\bbb
  ds^2&=&\Ht^{-1}\left(-d\ttil^2+d\xt_5^2+\Kt d\vt^2
	+2\At_i d\xt_i d\vt\right)+d\xt\cdot d\xt+dy\cdot dy\nonumber\\
  B_{\ut\vt}&=&-G_{\ut\vt}=\hf \Ht^{-1}\quad,\qquad
  B_{\vt i}=-G_{\vt i}=-\Ht^{-1}\,\At_i\quad,\qquad
	e^{-2\Phi}=\Ht\quad,
\eee
where $\Ht$, $\Kt$ are harmonic functions, $\vec y$
parametrizes the directions 6789, and 
the field strength $F=d\At$ satisfies $d^*F=0$,
appropriate to the source~\pref{sourceconfig}:
\be
  \Ht=\frac{n_5}{|\xt-\Ft|^2}\quad,\qquad
  \Kt=\frac{n_5|{\Ft'}|^2}{|\xt-\Ft|^2}\quad,\qquad
  \At=-\frac{n_5{\Ft'}_i}{|\xt-\Ft|^2}\quad.
\ee
In order to be able to simply U-dualize this source configuration
to the F1-NS5 duality frame, it is necessary to smear the
source along $\xt_5$.%
\footnote{The smearing operation is a linear operation on
harmonic functions, hence preserves the property of solving
the supergravity equations of motion.}
Defining
\be
  \zt=\xt_1+i\xt_2\quad,\qquad \wt=\xt_3+i\xt_4\quad,
\ee
one finds after averaging the source over $\xt_5$ that
\bbb
  \Ht&=&\frac{n_5}{f_0}		\nonumber\\
  \Kt&=&\frac{n_1}{f_0}		\nonumber\\
  \At_1&=&-\frac{\sqrt{n_1n_5}\; a \, y_2}{|\zt|^2}
	\frac{|\zt|^2+|\wt|^2+a^2-f_0}{f_0}	\\
  \At_2&=&\frac{\sqrt{n_1n_5}\; a \, y_1}{|\zt|^2}
	\frac{|\zt|^2+|\wt|^2+a^2-f_0}{f_0}	\nonumber\\
  f_0&=&\Bigl[(|\zt|^2+|\wt|^2+a^2)^2-4a^2|\zt|^2\Bigr]^{1/2}\ .
  	\nonumber
\label{smharm}
\eee
The solution is now ready for U-duality transformation
to the F1-NS5 frame; the result is~\cite{Lunin:2001fv,Lunin:2001jy}
\bbb
  ds^2&=&(HK)^{-1/2}\left[-(dt-A\cdot dx)^2+(dx_5+B\cdot dx)^2\right]
	+(HK)^{1/2}dx\cdot dx+dy\cdot dy\nonumber\\
  B_{ti}&=&K^{-1}B_i\quad,\qquad B_{i5}=-K^{-1}A_i\quad,
\eee
where the F1-NS5 harmonic functions have the same form
in terms of the dual (untilded) coordinates as the harmonics
of the P-F1 solution
$H(x)=\Ht(\xt)$, $K=\Kt$, $A_i=\At_i$; 
and $B_i$ is the dual gauge field of $A_i$ in
the directions 1234,
$dB=-^*dA$,
\bbb
  B_3&=&-\frac{\sqrt{n_1n_5}\; a \, y_4}{|\wt|^2}
        \frac{|\zt|^2+|\wt|^2-a^2-f_0}{f_0}\nonumber\\
  B_4&=&\frac{\sqrt{n_1n_5}\; a \, y_3}{|\wt|^2}
        \frac{|\zt|^2+|\wt|^2-a^2-f_0}{f_0}\ ;
\eee
for more details, the reader may consult~\cite{Lunin:2001fv,Lunin:2001jy}.

One may readily verify that for $a=\sqrt{n_1n_5}/N$ 
one recovers the $(AdS_3\times S^3)/\IZ_N$ orbifold geometry
\bbb
  ds^2/\rads^2&=&-(r^2+\coeff{a^2}{n_1n_5})\,dt^2
	+r^2\,dx_5^2 +\frac{dr^2}{r^2+\frac{a^2}{n_1n_5}} 
		\label{orbgeom} \\
	& &\qquad+ \Bigl(d\theta^2
		+\cos^2\theta(d\psi-\coeff{a}{\sqrt{n_1n_5}}dx_5)^2
		+\sin^2\theta(d\chi-\coeff{a}{\sqrt{n_1n_5}}dt)^2
	\Bigr)\nonumber
\eee
after the change of variables
\bbb
  z&=& \sqrt{n_1n_5 r^2+a^2}\,\sin\theta\,e^{i\chi}\nonumber\\
  w&=& \sqrt{n_1n_5}\;r\,\cos\theta\,e^{i\psi}
	\label{varch}\\
  x_5&=&\phi\ .\nonumber
\eee
The form of the metric in the subspace parametrized by
$r$, $\theta$, $x_5$, and $\psi$ near the singularity at 
$r=0$, $\theta=\pi/2$, is locally of the form $\IR^4/\IZ_N$:
\be
  ds_4^2\sim d(r/a)^2+d\theta^2+\frac1{N^2}(r/a)^2dx_5^2
  	+(\theta-\pi/2)^2(d\psi-\frac1N dx_5)^2
\ee
(a standard form of the ALE metric at the orbifold point results from the
substitution $r/a=\rho\,\sin\alpha$, $\theta-\pi/2=\rho\,\cos\alpha$).
So indeed, $(AdS_3\times S^3)/\IZ_N$ corresponds to the
dual source~\pref{sourceconfig}.

Since the dual string source consists of $N$ strands
which coincide in spacetime, 
the $N$-wound source can split into $N$ separate sources
at no cost in energy.
These $N$ strands each carry $n_5/N$ units of winding, and
a fraction $n_1/N$ of the momentum%
\footnote{N.B. It is not required that $N$ divides $n_1$
in the original F1-NS5 system; the quantization of the
background string charge is not visible in the formalism
of~\cite{Giveon:1998ns}.  For the purposes of the present
discussion, one can either distribute the momentum
approximately equally among the $N$ separated strands,
or assume that $N$ divides $n_1$ as well.}.
The $N$ separate strands of P-F1 source yield
a moduli space of multicenter solutions, obtained
by replacing the harmonic functions and gauge fields~\pref{smharm}
by their multicenter counterparts
\bbb
  \Ht&=&\frac{n_5}{N}\sum_{\alpha=1}^{N}\frac1{f_{0,\alpha}}\nonumber\\
  \Kt&=&\frac{n_1}{N}\sum_{\alpha=1}^{N}\frac1{f_{0,\alpha}}	
\label{multicent} \\
  \At_1&=&-\frac{\sqrt{n_1n_5}\; a}{N}\sum_{\alpha=1}^{N}
		\frac{(y_2-y_{2,\alpha})}{|\zt-\zt_\alpha|^2}
	\frac{|\zt-\zt_\alpha|^2+|\wt-\wt_\alpha|^2
		+a^2-f_{0,\alpha}}{f_{0,\alpha}}\nonumber\\
  \At_2&=&\frac{\sqrt{n_1n_5}\; a}{N}\sum_{\alpha=1}^{N} 
		\frac{(y_1-y_{1,\alpha})}{|\zt-\zt_\alpha|^2}
	\frac{|\zt-\zt_\alpha|^2+|\wt-\wt_\alpha|^2
		+a^2-f_{0,\alpha}}{f_{0,\alpha}}\nonumber\\
  f_{0,\alpha}&=&\Bigl[(|\zt-\zt_\alpha|^2+|\wt-\wt_\alpha|^2+a^2)^2
	-4a^2|\zt-\zt_\alpha|^2\Bigr]^{1/2}\ .	\nonumber
\eee
One can show that, expanding the metric near the sources
for separations $|z_\alpha|,|w_\alpha|\ll a$,
the metric assumes the ALE form
\bbb
  ds^2/\rads^2 &=& V^{-1}(\vec x)(dx_5+\vec\omega\cdot d\vec x)^2
              +V(\vec x)d\vec x\cdot d\vec x  \nonumber\\
  V(\vec x) &=& \frac{\sqrt{n_1n_5}}{Na}\sum_{i=1}^{N} |\vec x-\vec x_i|^{-1}
  	\quad ,\qquad
  	dV=\ ^*d\omega\ ;
\label{alemetric}
\eee
here $\vec x$ is the three-vector of coordinates
out of the directions 1234 that is locally transverse to
the ring source.  In other words, $V$ is the near-source
limit of $(HK)^{1/2}$, and $\vec\omega$ is the 
near-source limit of $\vec B$ in the appropriate three directions.
The fourth parameter of the multicenter solution is 
the separation of the sources from one another 
in the angular direction along the ring,
or equivalently in the T-dual coordinate $\xt_5$.
Upon smearing over $\xt_5$, these solutions separated
along this angular direction are no different
from the single center solutions~\pref{smharm}; however,
string dynamics on $(AdS_3\times S^3)/\IZ_N$
knows about this deformation -- for instance,
in the analogous situation of
string theory on $\IR^4/\IZ_N$, this deformation
amounts to adjusting the NS $B$-flux through the $N-1$
collapsed cycles of the ALE space, while maintaining
the conical metric of $\IR^4/\IZ_N$.
Thus, the $(AdS_3\times S^3)/\IZ_N$ orbifold spacetime yields
a variation on the well-known duality between
ALE spaces and NS fivebranes~\cite{Kutasov:1996te}.

The couplings to the twisted
sector marginal operators~\cite{Martinec:2001cf}
\be
  \Sigma_q^{\sst SL(2)}\Sigma_{N-q}^{\sst SU(2)}
  	\quad,\qquad q=1,\ldots,N-1
\label{twistops}
\ee
of the string theory orbifold $(AdS_3\times S^3)/\IZ_N$
should therefore be identified with the moduli deformations
of the multicenter solution
(the four overall translation modes decouple
in the near-horizon limit).
Indeed, in the flat space limit $k\to\infty$
these operators become twisted moduli of the $\IR^4/\IZ_N$
orbifold CFT.
The twist operators~\pref{twistops} lie in the $(2,2)$
representation of the global $SU(2)\times SU(2)$ R-symmetry,
\ie\ they are a vector in the $\IR^4$ 
transverse to the fivebranes (the operators~\pref{twistops}
are the bottom $(-,-)$ components of the multiplet).
The operators~\pref{twistops}
are products of parafermions from the $SL(2)$ and $SU(2)$
current algebra theories, and thus identical to the moduli of the 
$(\frac{SL(2,\IC)}{SU(2)\times \IR}\times\frac{SU(2)}{U(1)})/\IZ_N$
sigma model of~\cite{Giveon:1999px} describing a decoupling
limit of slightly separated fivebranes; 
in the latter context, the deformations
{\it are} the operators that move fivebranes in 
their transverse space (see~\cite{Giveon:1999px} 
and section~\ref{fivebranes} below).  
In the fivebrane sigma model, it is
known that at finite distance in the moduli space one
finds singular worldsheet theories corresponding to points where
two or more fivebranes come together.  
Therefore it is unreasonable that the fivebrane sources
should coincide as in figure~\orbAdS, since this should
yield a singular worldsheet CFT.

The point described by the coset model has the fivebranes arranged
in a $\IZ_N$ symmetric fashion along a two-dimensional plane.
This strongly suggests that we should interpret the magnetic $\IZ_N$
symmetry of the orbifold theory in terms of a $\IZ_N$
symmetric arrangement of the unsmeared sources
around the ring of figure~{\orbAdS},
see figure~{\altorbAdS}.
The separation of the $N$ strands at the orbifold
point is in a coordinate along which the source
has been smeared, so the deformation along that direction will
be invisible in the geometry,
just as in the B-field deformation of $\IR^4/\IZ_N$.

\begin{figure}[ht]
\begin{center}
\[
\mbox{\hskip -3cm
\begin{picture}(250,250)(0,0)
\includegraphics{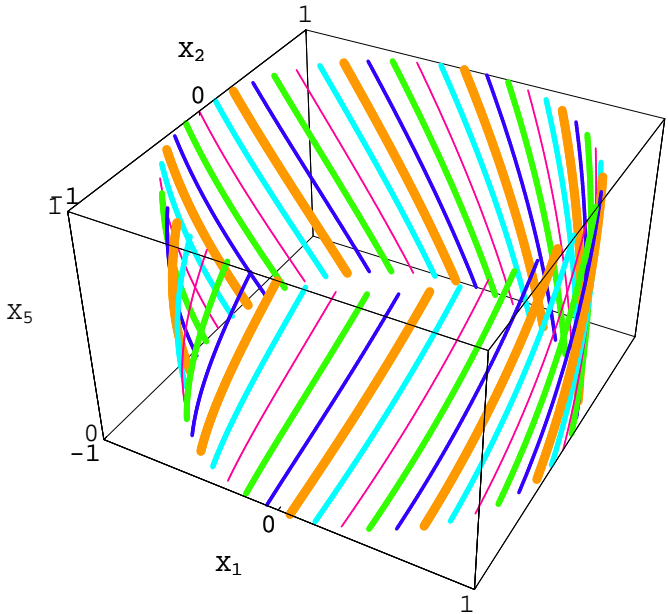}
\end{picture}} 
\]
\caption{\it
In the actual orbifold, the string source becomes $N$ strings
arrayed in a $\IZ_N$ symmetric fashion on a circle;
here $n_5=50$, and $N=5$.  The individual strings have
been depicted with
different colors and thicknesses for ease of visualization.
}
\end{center}
\end{figure}

In particular, we see that at finite distance
in the moduli space, where strands of the dual P-F1
strings come together, the worldsheet CFT becomes singular.
The point where all the strings come together 
is the configuration depicted in figure~{\orbAdS};
we now see that this configuration lies at a singular
point in the moduli space, where perturbative string
theory breaks down.
The fact that it takes finite energy above the $AdS$
vacuum to reach a singular CFT is consistent with
the fact that the strands of the dual string
are separated and have no moduli in the global case $N=1$ depicted
in figure~{\orbAdS}; 
it thus takes finite energy to push the fivebranes together.  
The scale of this energy $\LL_0\sim O(n_1)$
is consistent with the fact that the separation of the 
strands scales as $\sqrt{n_1}$ (see the discussion below
equation~\pref{strconfig}), and so it takes an energy
that scales as $n_1$ to push a pair of strands together. 

\vskip 0.5cm
\vbox{
\noindent{\sl Remarks:}

{\it 1.)}
It is important to note that, in the present context,
U-duality and orbifolding do not commute.  In the
original F1-NS5 background, the effect of the orbifold
is to transform the background from the $SL(2)$
invariant ground state of the spacetime CFT
to a particular excited state}
\be
  (\sigma_N^{--})^{n_1n_5/N}\ket{0}
\label{BPSstate}
\ee
on the BPS line $\LL_0+\frac{c}{24}=\TT_0^3$.%
\footnote{We use the notation of the BPS states
arising in the symmetric orbifold theory
$(T^4)^{n_1n_5}/S_{n_1n_5}$, as twist ground states
under the symmetric group.  More generally, we can characterize
the BPS states by the conjugacy class in the symmetric group
$S_{n_1n_5}$ that they correspond to, and this sets up an
unambiguous relation between the symmetric orbifold twist states
and the BPS states of the F1-NS5 background with
vanishing RR fields, the latter arising at a rather
different point in the moduli space of the spacetime CFT%
~\cite{Larsen:1999uk}.}
Carrying the $\IZ_N$ symmetry that one wants to quotient by
through the chain of dualities to the P-F1 frame,
one finds that it acts there as an axial $\IZ_N$
shift (since the $\xt_5$ circle is T-dual to the
$x_5$ circle of the F1-NS5 frame) together with
a $\IZ_N$ rotation in the $\xt_1$-$\xt_2$ plane.
Thus the result of such an orbifold in the P-F1
frame would produce a string sitting in a kind of Melvin geometry,
and not a transformed oscillator state in flat space,
as is suggested by the above analysis.

{\it 2.)}
A related issue is the 
suggestion in~\cite{Martinec:2001cf},
that there are new, nonperturbative kinds of orbifold constructions
for $N | n_1n_5$ but $N\!\!\!\not | \,n_5$.
There are U-duality transformations that take
$n_1$, $n_5$ to $n_1'$, $n_5'$ while keeping $n_1n_5=n_1'n_5'$,
and such that $N|n_5'$ but $N\!\!\!\not | \,n_5$%
~\cite{Larsen:1999uk};
the fact that there is a string theory orbifold in one
U-duality frame but not the other suggested that
there might be new kinds of orbifolds, hitherto unknown.
We now understand that there is a BPS state~\pref{BPSstate}
having the appropriate geometry $(AdS_3\times S^3)/\IZ_N$,
but it will not have any kind of standard orbifold construction,
in the sense of having twisted sectors of the usual type, \etc.

{\it 3.)}
Nevertheless, all P-F1 sources 
of the form~\pref{strconfig} U-dualize (after smearing) 
into BPS geometries in the F1-NS5 frame that seem to be 
exact worldsheet CFT's.  This is because all such geometries are
1/4 BPS, preserving eight supersymmetries.  This amount of supersymmetry
in spacetime is equivalent to $\NN=(4,4)$ supersymmetry on the worldsheet;
sigma models with this much supersymmetry are exact.  
In an orbifold CFT, the twisted sectors
are an essential ingredient describing string
theory on the geometry~\pref{orbgeom}, in particular they 
resolve the singularity (provided there are no strong
coupling singularities such as coincident fivebranes).
Presumably there are states similarly localized
near the singularity of the more general class of geometries
associated to~\pref{strconfig}, for instance in the conical
geometries of the previous remark; it would be very interesting to
understand them, for instance their asymptotic density
as discussed in~\cite{Harvey:2001wm}, and their role
in resolving the singularity in the geometry.

{\it 4.)}
The state that results from the orbifold operation is distinct
from the corresponding BPS state of the $\NN=4$ Liouville theory
associated to $AdS$ Chern-Simons supergravity,
which is also associated to conical defects.  As discussed
for example in~\cite{Martinec:1998wm},
the Liouville field is built (nonlocally) 
from the current sector of the spacetime CFT,
and its Hilbert space has BPS states 
with the same quantum numbers as~\pref{BPSstate};
however, these states are distinct.  In particular, there is
no way to distinguish different BPS states with the same quantum numbers 
$\JJ^3_0$, $\LL_0$ via their properties under the $\NN=4$ currents.
We will elaborate upon the role of the Liouville field below.

\vskip .5cm

To summarize,
the orbifold operation in perturbative string theory
maps global $AdS_3\times S^3$ to the geometry~\pref{orbgeom}
of $(AdS_3\times S^3)/\IZ_N$.  
We can then consider the image of this map under
the duality between bulk and boundary theories
under the AdS/CFT correspondence.
The global geometry $AdS_3\times S^3$ maps to the
$SL(2)$ invariant vacuum state of the dual CFT;
correspondingly, the orbifold geometry $(AdS_3\times S^3)/\IZ_N$
maps to the BPS state~\pref{BPSstate}.

This correspondence has a fascinating reflection in the 
orbifold (spacetime) CFT $(T^4)^{n_1n_5}/S_{n_1n_5}$,
which appears at a different point in the 
$\Gamma(n_1n_5)\backslash O(5,4)/O(5)\times O(4)$ moduli space
of the spacetime theory.%
\footnote{See for instance~\cite{Larsen:1999uk}
for a discussion of this moduli space, and in particular where 
the symmetric orbifold boundary theory $(T^4)^{n_1n_5}/S_{n_1n_5}$
and the perturbative bulk string theories
of~\cite{Giveon:1998ns} are situated in it.}
Twist field correlation functions of a free field orbifold CFT 
may be computed by a general procedure that employs
branched coverings of the Riemann surface on which
the CFT vertex operators are inserted%
~\cite{Dixon:1987qv,Hamidi:1987vh,Bershadsky:1987fv}.
The ramification points of the cover are the locations
of the twist operator insertions.  More recently,
Lunin and Mathur~\cite{Lunin:2000yv,Lunin:2001pw}
have adapted this method to compute the
correlation functions of the twist operators of
the symmetric orbifold $(T^4)^{n_1n_5}/S_{n_1n_5}$.
In this case, the ramification has the direct interpretation
of sewing together the different copies of $T^4$ 
in the spacetime CFT, and there are remarkable simplifications
in the structure of the correlation functions when one
works on the covering space.  For the particular case of
the twist ground state~\pref{BPSstate}, the two-point function
corresponding to the propagation of this state is 
computed by passing to an $N$-fold branched cover,
$w=z^{1/N}$, of the CFT parameter space.  
Apart from the transformation of the measure
(and a universal factor that implements the correct R-charge
of the twist in the case of superconformal field theory),
the covering space Riemann surface 
is the smooth two-sphere, and one is instructed to
compute the vacuum partition function on the cover.  
The central charge $\tilde c$ of
the covering space CFT is $1/N$ of that of the
original spacetime CFT, $c=N\tilde c$,
because of the $N$ copies of $T^4$ CFT that have been
sewn together in making the cover.%
\footnote{This is related to the fact that in the long
string sector of~\cite{Maldacena:1996ds} there is
an equivalence between a theory of central charge $6n_1n_5$
on a circle of radius $R$ and a theory of central charge $6$
on a circle of radius $n_1n_5 R$.}

But this structure is exactly that of
the perturbative string theory orbifold
$(AdS_3\times S^3)/\IZ_N$ -- namely, there is an $N$-fold covering 
space which is smooth, \ie\ the global $AdS_3\times S^3$ vacuum spacetime
(see figure~\znorb).
For example, the relation $c=N\tilde c$ of the central
charges was noted in~\cite{Martinec:1998wm}.
It is standard to identify the parameter space of the CFT with
the conformal boundary of AdS; the Euclidean continuation of $AdS_3$
to $H_3^+=SL(2,\IC)/SU(2)$ has conformal boundary $S^2$,
and the orbifold operation of the bulk theory on $H_3^+$
directly generates the branched cover seen 
in the representations of the twist correlators 
of the boundary theory.
Once again we see a rather precise correspondence between
bulk and boundary theories.  Of course, the perturbative
bulk orbifold and the symmetric orbifold boundary theory
are at different points in the moduli space of onebrane-fivebrane
backgrounds, so one cannot compare directly; however,
this remarkable parallel suggests to us that this property
of covering onto smooth global $AdS$ may be protected by supersymmetry
as one travels across the moduli space.

\begin{figure}[ht]
\begin{center}
\[
\mbox{\begin{picture}(263,179)(0,0)
\includegraphics{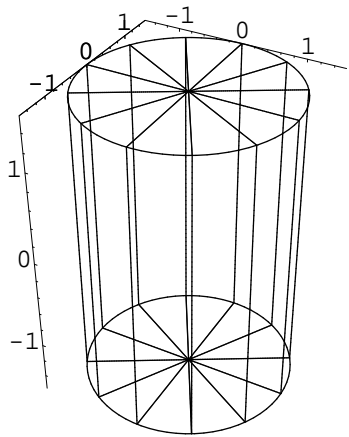}
\end{picture}} 
\]
\caption{\it
The $(AdS_3\times S^3)/\IZ_N$ orbifold geometry covers smoothly
onto global $AdS_3\times S^3$.  Here we have indicated
the slicing of the global space into fundamental domains,
with the quotient space being the conically singular geometry.
This same covering procedure is involved in computations
of twist operators in the $(T^4)^{n_1n_5}/S_{n_1n_5}$
orbifold related to the dual spacetime CFT.
}
\end{center}
\end{figure}

Although it would appear that the perturbative string orbifold
procedure can be directly mapped 
to a corresponding representation
in the exact boundary CFT dual, perhaps a note of caution
is in order -- the precise correspondence may be special to BPS states.
In~\cite{Martinec:1998wm}, a set of nonsupersymmetric orbifolds
$(AdS_3/\IZ_N)\times S^3$ was also considered.  In this case,
$N$ can be arbitrarily large (in particular it need not divide
$n_1n_5$).  The perturbative spectrum of such orbifolds is tachyonic,
indicating~\cite{Martinec:1998wm,Adams:2001sv}
that they decay via the emission of a pulse
of string excitations, eventually settling down to an ordinary
excited state of $AdS_3\times S^3$ with the same 
quantum numbers.  This state cannot have 
a naive covering space interpretation
of the type considered above -- the covering CFT would have 
central charge $\tilde c=6n_1n_5/N$, and $N$ can be arbitrarily large.
Of course, the effect on the central charge is a special
construction for twist operators of the symmetric group.
There is certainly a covering interpretation for general
$\IZ_N$ twists in terms of a CFT of the {\it same}
central charge as the original theory%
~\cite{Dixon:1987qv,Hamidi:1987vh,Bershadsky:1987fv},
but the path integral on the covering
space is constrained -- the fields must take on
$\IZ_N$ related values at $\IZ_N$ related points 
on the covering space.  It is only in the case of
the symmetric orbifold twist operators that the 
covering space path integral is an unconstrained path integral%
~\cite{Lunin:2000yv,Lunin:2001pw} in terms of
a theory with central charge $N$ times smaller
(\ie\ $n_1n_5/N$ copies of the component CFT
underlying the symmetric product).
There might be a description of 
the $(AdS_3/\IZ_N)\times S^3$ orbifold in terms of
a twist operator in the dual CFT, of this more
general type.

This description of BPS states and correlators 
in the spacetime CFT also sheds light on the role of
the Liouville field that summarizes Chern-Simons
$AdS_3$ supergravity~\cite{Coussaert:1995zp}.
The covering transformation yields a smooth Riemann surface
from one with $\IZ_N$ conical defects, but the corresponding
coordinate map $w=z^{1/N}$ affects the path integral
via the conformal anomaly.  As discussed 
for instance in~\cite{Martinec:1998wm}, the Liouville field
$\varphi_L=\log|\partial z/\partial w|^2$
summarizes the physical data of the Chern-Simons supergravity
connection, in this instance the coordinate map relating flat space
and the conical defect.
Correspondingly, in~\cite{Lunin:2000yv,Lunin:2001pw}
this same Liouville field arises in the covering onto
the vacuum spacetime via the conformal anomaly,
and for the same reason -- the gauge invariant modes
of Chern-Simons supergravity can be identified 
with the modes of the current sector of the spacetime CFT.


\section{\label{excitations} The excitation spectrum}

Having identified the state in the exact spacetime CFT
corresponding to the perturbative $(AdS_3\times S^3)/\IZ_N$ orbifold,
we can now give a direct interpretation of the spectrum
of low-energy perturbative string excitations around the orbifold.  
These consist of
\begin{itemize}
\item {\it Untwisted short string states.}
These are characterized by their representation labels
$(j,m,\bar m;j',m',\bar m')$ under the worldsheet
left and right $SL(2,\IR)\times SU(2)$ current algebra,
together with the oscillator content.  
The $SL(2)$ spin $j$ lies in the principal discrete series,
$j\in \IR$, $\hf\le j<(k-1)/2$.
For massless (supergravity) states, the orbifold restricts
roughly to states with $(m-\bar m)-(m'-\bar m')\in N\IZ$
(up to shifts due to the helicity content of the polarization state).
In particular, excitations with $m=\bar m$, $m'=\bar m'$
are unaffected.
\item {\it Twisted short string states.}
These are roughly spanned by acting with twisted oscillators,
and exponentials of the bosonized $J^3$, $J^{\prime 3}$ currents,
on the $SL(2,\IR)\times SU(2)$ parafermion primaries~\pref{twistops}.
In the untwisted theory, the gap to the first oscillator 
excitations came from solving the mass shell condition
\be
  \frac{-j(j-1)}{n_5}+\frac{j'(j'+1)}{n_5} +\N_\osc=\half\ ,
\ee
so the integer gap in oscillator number results in
a gap in spacetime energy $\tLL_0$ 
of order $\delta j\sim \sqrt n_5$ 
in the measuring conventions of the covering space. 
In the orbifold theory, the fractional moding of
twisted oscillators reduces
the gap in the oscillator spectrum by a factor $N$
(again, as measured on the covering space),
correspondingly the gap to the lowest stringy modes
becomes $\delta\tLL_0=\delta j\sim \sqrt{n_5}/N$.
In particular, for $N\sim n_5$, the gap becomes of the same order
as that of the supergravity modes -- there is no regime where supergravity
is an effective approximation to the dynamics!
\item {\it Long string states.}
These states arise from principal continuous series 
representations~\cite{Maldacena:2000hw} with $j=\hf+is$, and have energy
\be
  \tLL_0=\frac{n_5w}{4}+\frac1w\Bigl(\frac{1+4s^2}{4n_5}
  		+h_{\rm int}-\half\bigr)\ .
\ee
They describe strings that have escaped or are joining the background
ensemble.  Their coordinates are U-dual analogues
of the twisted sector moduli separating fivebranes
from the background ensemble.
In the orbifold theory, fractional spectral flow by $w=p/N$ 
is allowed, provided one flows by an amount $-p/N$ in $SU(2)$;
this means that the internal level is
\be
  h_{\rm int}=\frac{j'(j'+1)}{4n_5}-wm'+\frac{n_5}{4}w^2+\N_\osc\ .
\ee
Setting $m'=0$  for simplicity, one sees that
(after exciting a fermion oscillator $\N_\osc=\hf$ to satisfy 
the GSO projection) that there are long string states at 
(covering space) energy
\be
  \tLL_0=\frac{n_5}{N}\, p+\frac{N}{pn_5}\Bigl(\frac{1+4s^2}{4n_5}
  	+\frac{j'(j'+1)}{4n_5}\Bigr)\ .
\ee
The gap to the emergence of long string states is thus also reduced by
a factor of $N$; in particular, it is of the same order as
the gap in supergravity modes for $N\sim n_5$.

\end{itemize}

The energy scales of these various excitations undergoes
a {\it further} rescaling related to the orbifold identification.
As discussed in \cite{Martinec:2001cf}, 
the time coordinate $t$ in terms of which the metric takes an
asymptotic $AdS$ form, is rescaled by a factor of $N$ relative to
the time coordinate $\ttil$ of the covering space
(unorbifolded global $AdS$), $t=N\ttil$.
Hence all energy scales $\LL_0$ of the untwisted sector 
are rescaled by a factor of $1/N$ with respect to
the covering space energy $\tLL_0$,
\be
  \LL_0=\coeff1N\tLL_0\ .
\ee
In particular, the gap in supergravity excitations is 
the inverse $AdS$ radius $\rads^{-1}$ in global $AdS$,
and $(N\rads)^{-1}$ in the orbifold theory.  
In global $AdS$, the gap in the spacetime energy $m+\bar m$ and
angular momentum $m-\bar m$ of the $AdS$ modes of a given
supergravity field (fixing also the quantum numbers on $S^3$)
is accounted for by the fact that the modes are spanned by
the action of $\LL_{-1}$, $\bar\LL_{-1}$ on the highest
weight state:  
\be
  \Phi_{j\,m\,\bar m}^{\sst SL(2)}=(\LL_{-1})^{m}(\bar\LL_{-1})^{\bar m}
  	\Phi_{j\,0\,0}^{\sst SL(2)}\ .
\ee
Thus the gap in energy is $1/\rads$
(recall that spacetime energy is $\rads E=\LL_0+\bar\LL_0$).
In the orbifold spacetime, $\LL_{-1}$ descends from 
$\tilde\LL_{-N}$ on the covering space,
and does not span the full set of modes; thus, 
the $\Phi_{j\,m\,\bar m}^{\sst SL(2)}$ are {\it not} related
by the spacetime superconformal algebra.  
Shifting $m+\bar m$ by two (keeping $m-\bar m$ fixed)
results in a change of the energy by an amount $2/N\rads$.

This energy gap is inversely related to the return time 
of supergravity probes sent radially inward from the top 
of the throat in the full asymptotically flat D1-D5 geometry, 
\be
  \Delta t=\pi\frac{\sqrt{n_1n_5}}{a}\ ,
\ee
as noted in \cite{Lunin:2001jy}; for the
orbifold backgrounds, one has $\Delta t=N\pi$.
Our point here is that this time
scale remains visible in the near-horizon limit, as the
gap in the spectrum relative to the AdS scale.  The effect
of the orbifold is indeed to make a deeper gravitational
well in the center of AdS, as expected from the fact that
the conical defect starts to approach the black hole
limit from below at large $N$.  The closer one approaches
the black hole threshold, the smaller the gap in the
spectrum, due to the increasing redshift generated
by the large mass at the center of $AdS$.  
At the extremal black hole point $\LL_0=c/12=\TT_0^3$,
the spectrum becomes continuous (in the classical theory).

We may interpret these various excitations in terms of
the underlying brane dynamics as follows.
The untwisted string states are related to the singlet sector --
the overall $U(1)$ of the $U(N)$ gauge dynamics -- of the 
fivebrane dual.  This is in accord with the standard
AdS/CFT duality between gauge invariant local operators
and supergravity modes.  The twisted sectors are the `Cartan'
multiplets describing individual groups of $n_5/N$ fivebranes
(so one has the full set of Cartan modes for $N=n_5$),
since the twisted sector moduli describe relative overall
motion of the branes.  
The off-diagonal modes of fivebrane gauge dynamics
are described as the fractional D-strings of the orbifold.  
Indeed, the separation of strands in the dual string source
of global $AdS$ of figure~\globalAdS,
$\sqrt{n_1/n_5}\,\lstr$, 
corresponds (as it should) to the energy scale
of D-branes in the F1-NS5 frame, since 
$\gstr^{-1}=\sqrt{n_1/n_5}$.
In the orbifold theory, the separation
of the strands of the dual string decreases by a factor of $N$,
reducing the energy cost of fractional D-strings 
in the F1-NS5 frame by a corresponding factor.
Note that there are no noncompact spatial
directions into which the RR flux of these fractional branes may
escape; thus the fractional branes must always organize
themselves into regular representation branes
carrying no net fractional D1 charge; such branes can
then disappear by leaving the fivebranes and decaying into
untwisted string modes.  This is a reflection of Gauss' law
in the fivebrane gauge theory. 

One should also note that the twisted moduli of the orbifold
are effectively 1+1 dimensional scalar fields,
and so it is somewhat misleading to consider them as having
expectation values as we have been doing.  This is an
artifact of the tree level approximation to the orbifold
dynamics in perturbative string theory.  
At one loop, one will encounter an infrared
divergence arising from the fluctuations of the moduli,
and one will have to quantize them as 
collective coordinates (\cf~\cite{Fischler:1995tk}).
The moduli are thus the coordinates of a wavefunction for
the fivebrane sources, whose understanding will require
knowledge of the dynamics near the singular regions where
fivebranes coincide.

It was noted in~\cite{Martinec:2001cf} that the gap
to black hole states above the $(AdS_3\times S^3)/\IZ_N$
orbifold is of order $n_1n_5/N^2$, which is
of order $n_1/n_5=g_6^2$ when $N=n_5$%
\footnote{$N$ is required to $n_5$ in order that the orbifold
is not anomalous.}.
There are two interesting interpretations of this scale.
First, the energy to remove a fundamental string from the
background ensemble is of order $n_5$; thus, the black hole 
threshold is of order the energy required to remove all the
strings from the background and put them on their Coulomb branch.
Left behind would be a strong-coupling throat of the $n_5$ 
fivebranes, indicating the appearance of strong-coupling dynamics.
Second, the energy cost of D-branes is of order $1/g_6=\sqrt{n_1/n_5}$.%
\footnote{This cannot be reduced by a factor $N$ by the consideration
of fractional brane representations because, as mentioned above,
only regular representation branes are allowed.}
The BTZ threshold is the energy cost of $n_D\sim\sqrt{n_1/n_5}$ 
D-branes, so that the effective gauge coupling $g_6n_D\sim 1$,
and again the dynamics is strongly coupled.  
We regard these two estimates as yet another indication 
that the onset of black hole physics is a strong coupling phenomenon
from the point of view of string dynamics.


\section{\label{fivebranes}An aside on fivebranes}

The smearing of sources to obtain the geometry seen by
low-energy string theory has applications to other backgrounds,
in particular that of NS5-branes on the Coulomb branch
studied in~\cite{Giveon:1999px,Giveon:1999tq}.
There, it was claimed that the nonsingular coset CFT
$(\frac{SL(2,\IC)}{SU(2)\times \IR}\times\frac{SU(2)}{U(1)})/\IZ_{n_5}$%
\footnote{The Euclidean continuation of $SL(2,\IR)$ is
the hyperbolic space $H_3^+=SL(2,\IC)/SU(2)$.
The $\IZ_{n_5}$ identification is related to the GSO projection.}
describes the four-dimensional transverse space of separated fivebranes.  
On the other hand, supergravity would suggest
that the target space is that of the multicenter CHS throat solution
(directions along the fivebranes are suppressed)
\bbb
  ds^2 &=& -dt^2+H_5\Bigl(|dz|^2+|dw|^2\Bigr)\nonumber\\
  e^{2\Phi}&=&H_5\\
  dB&=&^*d\Phi\nonumber\\
  H_5&=&\sum_{\alpha=1}^{n_5}\frac{1}{|z-z_\alpha|^2+|w-w_\alpha|^2}\ .
  	\nonumber
\eee
To see the relation between these two geometries,
place the fivebranes at $z_\alpha=a\,\omega^\alpha$, $w_\alpha=0$
(where $\omega^{n_5}=1$), and smear the fields over the argument of $z$.
One finds that $H_5$ is the same harmonic function 
$\Ht$ of~\pref{multicent}.
After the change of variables~\pref{varch}, and defining
$r/a=\sinh(\rho)$, the resulting geometry is
\bbb
  ds^2&=& -dt^2+d\rho^2+d\theta^2
	+\frac{\sinh^2\rho\,\cos^2\theta}{\sinh^2\rho+\cos^2\theta}
		d\psi^2
	+\frac{\cosh^2\rho\,\sin^2\theta}{\sinh^2\rho+\cos^2\theta}
		d\phi^2\nonumber\\
  \Phi&=& \log[\sinh^2\rho+\cos^2\theta]\ .
\label{coset}
\eee
Remarkably, this {\it is} the gometry of a coset CFT.
A general class of Lorentzian coset models of the form
$\frac{SL(2,\IR)\times SU(2)\times U(1)}{U(1)\times U(1)}$,
describing charged black holes,
was considered in~\cite{Gershon:1994xr}.
After a Wick rotation to Euclidean $AdS_3$,
and taking the limit of the charge parameter $Q^2\to -1$ 
(in the conventions of~\cite{Gershon:1994xr}), 
one finds that the extra $U(1)$ decouples
and the geometry becomes precisely that of~\pref{coset}!
Thus, smearing of the source provides the connection
between the CHS geometry and the exact sigma model
description of~\cite{Giveon:1999px}.  Note that smearing does
not remove the singularities in the metric and dilaton;
these still appear in~\pref{coset}, however they are apparently
harmless in string theory, since the correlation functions 
of the coset are entirely well-behaved (at least 
at sufficiently low energy~\cite{Giveon:1999px,Giveon:1999tq}).
This is a reflection of the absence of throat dynamics
for separated fivebranes.
Once again, the string theory remembers that the sources are localized
and not smeared; for instance, D-strings stretch between
specific $\IZ_N$ symmetric points along the locus 
$\sinh^2\rho+\cos^2\theta=0$~\cite{Maldacena:2001ky,Elitzur:2000pq}.


\section{\label{btzsection}Comments on BTZ orbifolds, and cosmology}

In this section we try to apply some of the lessons learned above
in the study of rotational orbifolds of $AdS_3\times S^3$
to other contexts, specifically the class of orbifolds that
realize BTZ black holes as quotients of $AdS_3$%
~\cite{Banados:1992wn,Banados:1993gq}.

\subsection{The BTZ geometry as an $AdS_3$ quotient}

Group elements $h\in SL(2,\IR)$ lie in one of
three conjugacy classes -- elliptic, parabolic, and hyperbolic,
depending on whether $|\Tr[h]|$ is less than, equal to, or
greater than two, respectively (the trace is in
the two-dimensional representation).  Elliptic elements are
conjugate to a rotation, and lead to the class of conical
defect spacetimes discussed above.  The orbifold
\be
  g\sim hgh\quad,\qquad g\in SL(2,\IR)
\label{btzorb}
\ee
of $AdS_3$ by the action of the discrete group generated
by a hyperbolic element $h$, leads to a fundamental domain
of the identification which contains the BTZ black hole spacetime with 
zero angular momentum (\(J=0\))%
~\cite{Banados:1992wn,Banados:1993gq}
(the identification by a parabolic element
yields the extremal (\(\rads M=J\)) BTZ black hole).
Let us parametrize $g\in SL(2,\IR)$ via the analogue of Euler angles
\bbb
  g&=&e^{(t-\phi)i\sigma_2/2}\;e^{\rho\sigma_3}\;e^{(t+\phi)i\sigma_2/2}
	\equiv\pmatrix{a&b\cr c&d} 
\label{eulerang} \\
   &=&\pmatrix{ \cos (t)\,\cosh (\rho) + \cos (\phi)\,\sinh (\rho) & 
	\sin(t)\,\cosh (\rho) + \sin (\phi)\,\sinh (\rho) \cr 
  - \sin (t) \, \cosh (\rho) + \sin (\phi)\,\sinh (\rho) & 
	\cos (t)\,\cosh (\rho) - \cos (\phi)\,\sinh (\rho) \cr  }\quad;
	\nonumber
\eee
in these coordinates, the metric on $AdS_3$ takes the form
\be
  ds_{\sst AdS}^2=\rads^2
	\Bigl(-\cosh^2\rho\,dt^2+d\rho^2+\sinh^2\rho\,d\phi^2\Bigr)\ .
\label{globalmetric}
\ee

\begin{figure}[ht]
\begin{center}
\[
\mbox{\begin{picture}(300,300)(0,0)
\includegraphics{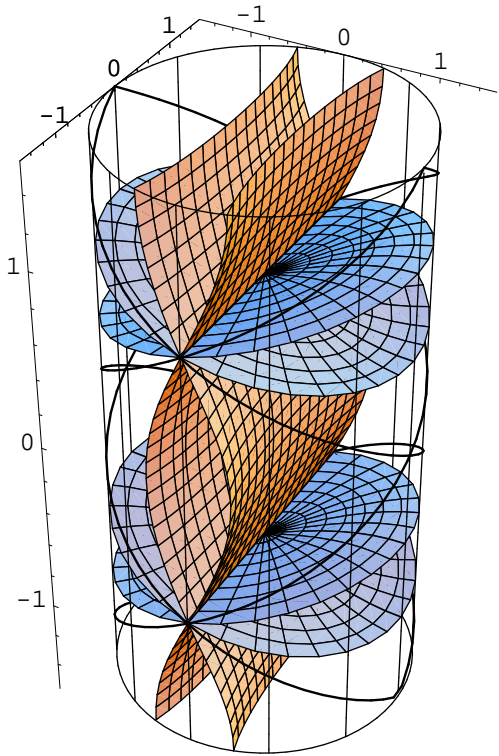}
\end{picture}} 
\]
\caption{\it
The fundamental domain of the identification of $AdS_3$
corresponding to a BTZ black hole.
The vertical (amber) wavy surfaces are identified by
a spacelike translation; also the
horizontal (blue) disks are identified by
a timelike translation.  
The identification has surfaces of null identification
which intersect the boundary along the thick (black) helices.
The intersection of all these surfaces is the spacelike
BTZ black hole singularity.
}
\end{center}
\end{figure}

The BTZ black hole geometry results from 
the identification~\pref{btzorb}, with
$h=\exp[{\pi\rp}\sigma_3]$; the mass of the BTZ black hole
will then be $\ell M=\txfc{1}{2}n_1 n_5\,\rp^{2}$, with $\rp$ the horizon 
radius in units of the $AdS$ curvature scale $\rads$. 
The fundamental domain of this orbifold is depicted in
figure~\alttrueBTZregion, with the radial coordinate redefined as
$\sinh[\rho]=\tan[\theta]$ in order to view $AdS_3$ as
the canonical solid cylinder. 
Slices through the $AdS$ cylinder at $\sin\phi=0$ and
$\cos\phi=0$ are depicted in figure~\BTZslices;
the slice $\sin\phi=0$ exhibits the identification,
while the slice $\cos\phi=0$ reveals the standard Penrose
diagram of the Kruskal extension of the black hole geometry.

\begin{figure}[ht]
\begin{center}
\[
\mbox{\begin{picture}(300,250)(50,40)
\includegraphics{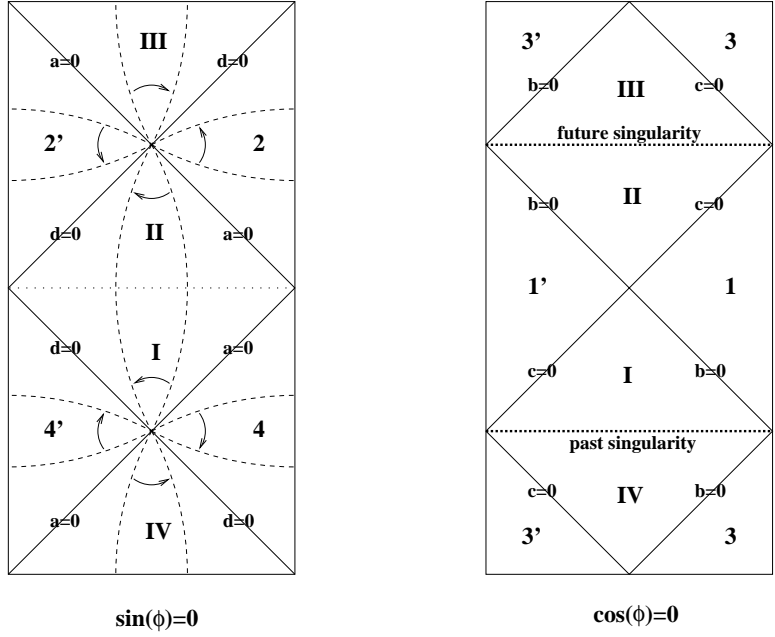}
\end{picture}} 
\]
\caption{\it
Slices of the geometry of figure~\alttrueBTZregion, at
$\sin\phi=0$ and at $\cos\phi=0$.  The slice $\sin\phi=0$
exhibits the identification of global $AdS$ involved,
while the slice $\cos\phi=0$ shows the Penrose diagram.
The labels of the various regions 
are adopted from the conventions of~\cite{Elitzur:2002rt}.
Also, the $2\times 2$ matrix element of $g$ which vanishes
along various null curves is indicated.
}
\end{center}
\end{figure}

For completeness, we also give the metric geometry
that results from the quotient, and its relation
to the group manifold, following%
~\cite{Banados:1992wn,Banados:1993gq}.
In the standard BTZ coordinates, the black hole metric takes the form
\be
  ds_\btz^2/\rads^2=-({\rhat^2-\rp^2})\,d\that^2
  	+({\rhat^2-\rp^2})^{-1}\,d\rhat^2
	+\rhat^2d\phihat^2\ ,
\label{btzmetric}
\ee
with $\phihat\sim\phihat+2\pi$.
The relation to the global coordinates~\pref{globalmetric}
is best established in the Kruskal extension of~\pref{btzmetric}
\be
  u=\sqrt{\frac{\rhat-\rp}{\rhat+\rp}}\;\exp[+\rp\that]
	\quad,\qquad
  v=\sqrt{\frac{\rhat-\rp}{\rhat+\rp}}\;\exp[-\rp\that]
	\ ,
\ee
for which the metric is
\be
  ds^2_\btz/\rads^2=(1-uv)^{-2}\bigl[4\,dudv 
  	+ \rp^2(1+uv)^2d\phihat^2\bigr]\ .
\label{kruskalmetric}
\ee
The black hole horizon is the surface $uv=0$,
the singularities are at $uv=-1$, and the asymptotically
locally $AdS$ boundary is at $uv=+1$.
Near the singularity, the geometry is like that of a
Milne universe as the $\phihat$ circle shrinks to zero size.
In these coordinates, the $SL(2,\IR)$ matrix 
$g=\left({a~b\atop c~d}\right)$ of~\pref{eulerang} is written
\bbb
  a &=& \Bigl(\frac{\rhat}{\rp}\Bigr)\,\exp[+\rp\phihat]
  	= \Bigl(\frac{1+uv}{1-uv}\Bigr)\,\exp[+\rp\phihat] 
  	\nonumber\\
  b &=& \Bigl({\frac{\rhat^2}{\rp^2}-1}\Bigr)^{1/2}\;\exp[+\rp\that]
  	= \frac{2u}{1-uv} \nonumber\\
  c &=& \Bigl({\frac{\rhat^2}{\rp^2}-1}\Bigr)^{1/2}\;\exp[-\rp\that]
  	= \frac{2v}{1-uv} \\
  d &=& \Bigl(\frac{\rhat}{\rp}\Bigr)\,\exp[-\rp\phihat]
  	= \Bigl(\frac{1+uv}{1-uv}\Bigr)\,\exp[-\rp\phihat] 
  	\nonumber
\eee
(recall the identification is $\phihat\sim\phihat+2\pi$).
This parametrization covers regions $1,1',I,II$ in figure~\BTZslices;
the same parametrization covers regions $3,3',III,IV$ if we 
send $g\to -g$.  The regions $2,2'$ are covered by
\bbb
  a &=& \sinh[\rho']\,e^{t'}\quad,\qquad
  b = \cosh[\rho']\,e^{\phi'} \nonumber\\
  c &=& -\cosh[\rho']\,e^{-\phi'} \quad,\qquad
  d =  -\sinh[\rho']\,e^{-t'} 
\eee
(and similarly regions $4,4'$ by sending $g\to -g$);
the metric in these regions is 
\be 
  ds^2 = \rads^2\Bigl(
	-\sinh^2\rho'\,d{t'}^2+ d{\rho'}^2+\cosh^2\rho'\,d{\phi'}^2\Bigr)\ .
\ee
The action of the identification in these regions is
$t'\sim t'+2\pi\rp$, and thus has a singularity at $\rho'=0$.
This metric is identical to~\pref{btzmetric}
(via $\rhat/\rp=\cosh\rho'$, $\rp\that=t'$, $\rp\phihat=\phi'$),
except that the temporal coordinate is periodically identified
rather than the spatial coordinate.


\subsection{Is there an exact orbifold description?}

There have been several attempts to give meaning
to perturbative string theory on~\pref{btzorb}
and similar time-dependent orbifolds%
~\cite{%
Satoh:1997xf,%
Satoh:1998xe,%
Behrndt:1998wg,%
Behrndt:1999jp,%
Hemming:2001we,%
Cornalba:2002fi,%
Nekrasov:2002kf,%
Simon:2002ma,%
Tolley:2002cv,%
Balasubramanian:2002ry,%
Liu:2002ft,%
Elitzur:2002rt,%
Craps:2002ii%
}
Certainly, if string perturbation theory is sensible
on this orbifold one will reap valuable information
about the AdS/CFT correspondence.  Furthermore, 
near the singularity the geometry is locally that
of identification of Minkowski space under a discrete
boost -- the Milne spacetime -- that has figured in
several recent discussions of string cosmology%
~\cite{%
Horowitz:1991ap,%
Khoury:2001bz,%
Seiberg:2002hr,%
Nekrasov:2002kf,%
Simon:2002ma,%
Tolley:2002cv,%
Liu:2002ft,%
Craps:2002ii%
},
and is a prototypical example of a pre-big bang cosmological scenario
(\cf~\cite{Veneziano:2000pz} for a review).
In addition to the standard Kruskal extension of the
black hole, which is the fundamental domain of the region
where the identification is spacelike, there are
also regions of timelike identification, 
see figure~\alttrueBTZregion.

Naively, the effect of the orbifold procedure might
be expected to be the identification of the parameter
space cylinder of the spacetime CFT by the action of~\pref{btzorb}
on the corresponding conformal boundary of $SL(2,\IR)$.
If so, then the problem would reduce to understanding 2d field theory
in the cyclic Milne-like cosmological
spacetime in which the dual `boundary' CFT lives.
On the boundary, the BTZ and global coordinates
are related by
\be
  \exp[\pm\rp(\that\pm\phihat)] = 
	\pm\tan[(t\pm\phi)/2] \ ,
\label{boundcoords}
\ee
and the identification is $\phihat\sim\phihat+2\pi$.%
\footnote{It is worth noting that the boundary structure
of the identification is little changed upon incorporation of
$AdS_3$ angular momentum in the geometry.
The argument of the exponential on the
LHS of~\pref{boundcoords} is simply modified to
$(\that\pm\phihat)(\rp\mp\rminus)$, 
even though in the interior there is a rather dramatic 
modification of the Penrose diagram and causal structure,
(\cf~\cite{Banados:1992wn,Banados:1993gq}) --
an inner horizon develops, and the singularity
becomes timelike rather than spacelike.}

\begin{figure}[ht]
\begin{center}
\[
\mbox{\begin{picture}(200,200)(0,0)
\includegraphics{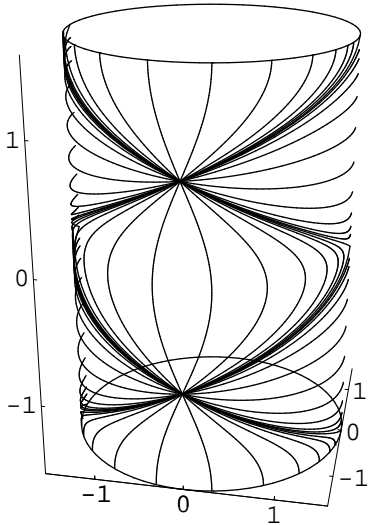}
\end{picture}} 
\]
\caption{\it
Covering of the BTZ spacetime onto global $AdS_3$.
The curves where the boundaries of the fundamental
domain intersect the boundary of $AdS_3$ are indicated
for half of the boundary.  The points of intersection of all
these curves ought to be the locations of twist operators
in the spacetime CFT.
}
\end{center}
\end{figure}

We have seen that the effect of a quotient of the bulk theory
by an elliptic $\IZ_N$ transformation 
amounts to consideration of a different state~\pref{BPSstate} 
in the dual CFT, whose correlators are indeed computed
by a $\IZ_N$ covering of the Riemann surface
on which they are defined, see figure~\znorb.  
For the orbifold~\pref{btzorb} related to the BTZ geometry,
the slicing of global $AdS_3$ into fundamental domains
under the identification is depicted in figure~\crunch.

The idea we wish to entertain 
is that the Milne-type BTZ singularity
is described in the exact dual spacetime CFT
as a `vertex operator' that lives at and implements
the big crunch/big bang on the boundary.
This picture is somewhat different 
from that of~\cite{Maldacena:2001kr},
where it was proposed to describe the black hole 
as a correlated state in the boundary theory (see also%
~\cite{Horowitz:1998xk,%
Balasubramanian:1998de,%
CarneirodaCunha:2001jf} 
for related discussions).
These descriptions employ the BTZ time of the static
asymptotic observer and therefore can't see
the big crunch/big bang singularity, 
since this occurs at infinite time in those coordinates.
The orbifold description is an attempt to
maintain the use of the global time, which is
more akin to the proper time of an infalling observer,
especially near the singularity.

The identification $\phihat\sim\phihat+2\pi$ of the BTZ orbifold
implies that the corresponding CFT state
is not the CFT vacuum.%
\footnote{Although apparently absent this identification
the appropriate state {\it is} the CFT vacuum;
for a discussion in a closely related context
of vacua related to the coordinate 
transformation~\pref{boundcoords}, 
see for instance~\cite{Spradlin:1999bn}.}
The appropriate state in the boundary CFT is the one
that yields the proper BTZ correlators, for instance the
two-point function~\cite{Keski-Vakkuri:1998nw,Louko:2000tp,Maldacena:2001kr}
\bbb
  \lefteqn{\hspace{-1cm}\vev{\OO(\xhat_+,\xhat_-)
	\OO(\xhat_+',\xhat_-')} \sim}
  \nonumber \\[0.2cm]
	& & \sum_{m=-\infty}^\infty
	\Bigl[\sinh[\coeff{\rp}{2}(\xhat_+ - \xhat_+'+2\pi m)]\;
	\sinh(\coeff{\rp}{2}(\xhat_- - \xhat_-'+2\pi m)]\Bigr]^{-2h}
\eee
between operators on the same Kruskal boundary;
here $\xhat_\pm=\that\pm\phihat$ and $h$ is the conformal weight.
As usual, the thermal nature of this state 
with respect to static observers in the (hatted) black hole
coordinates is reflected in the invariance of this expression 
under the transformation 
$\that\to\that+{2\pi i}/{\rp}$, which is an invariance
of the global coordinates, \cf~\pref{boundcoords}.
Mapping back to the global coordinates via~\pref{boundcoords} yields
(with similarly $x_\pm=t\pm\phi$)
\be
\vev{\OO(x_+,x_-)\OO(x_+',x_-')} \sim 
\sum_{m=-\infty}^\infty \ 
\left[f_{m}(x_+,x_+')\,\times\,f_{m}(x_-,x_-')\right]^{-h} \ ,
\label{globtwopt}
\ee
where \(f_{m}(x,x')\) has the form
\be
f_{m}(x,x')\,=\,(\,e^{m\pi\rp}\,\sin^2 \coeff{x}2\cos^2 \coeff{x'}2 +
e^{-m\pi\rp}\,\cos^2 \coeff{x}2\sin^2 \coeff{x'}2-\hf \sin x\sin x'\,) \ .
\ee
The expression in brackets in the $m=0$ term of this sum
reduces to the form $\sin^2[\hf(x_{+}-x'_{+})]\,\sin^2[\hf(x_{-}-x'_{-})]$,
which is the appropriate correlation function of scaling
operators (in the global coordinates) in the $SL(2,\IR)$ invariant vacuum;
the remaining terms may then be interpreted as coming from
the identification that yields the BTZ state.
Note that, as discussed above for the case of rotational orbifolds,
the vacuum path integral on the covering space is
not the right result -- the fields take on the same
values at points related by the group identification.%
\footnote{And thus we see that (as remarked above)
if it were not for this additional
restriction, the eternal black hole and the global $AdS$ vacuum
would be the same state, \cf~\cite{Spradlin:1999bn}.}
The relation of 1+1d scalar field modes respecting
identification under a boost, and the standard global modes
for a free scalar field, has been explored in%
~\cite{Tolley:2002cv} (see also~\cite{Liu:2002ft}). 

The analogue of the two-point function~\pref{globtwopt} for
dimension (1,0) fields can be used to compute the stress tensor 
of the black hole state, by taking the coincidence
limit and subtracting the pole term; one finds
(up to an additive constant)
\be
  T_{++}\sim \Bigl(2\sum_{m=1}^\infty \frac{1}{\sinh^2 m\pi \rp}\Bigr)
	\,\frac{1}{\sin^2 x_+}\quad ,
\ee
indicating that the energy of the black hole state is infinite
due to a rather bad divergence of the stress-energy along the
light cones emananating from the singular point of the identification.
This result is consistent with the known divergence of the bulk 
stress tensor at the fixed point of the identification or 
on the inner horizon of the spinning BTZ black hole~\cite{Steif:1994zv}. 
This strongly divergent stress-energy may be an indication that
the orbifold interpretation of the BTZ state is not viable.


There are regions of timelike identification 
in figure~\alttrueBTZregion.
The large radius boundary of the fundamental domain 
of the identification consists of two spacelike 
and two timelike cylinders
which meet at the points where the BTZ singularity intersects
the $AdS$ boundary; the spacelike and timelike cylinders
on the boundary are roughly speaking interchanged by the operation 
$t\leftrightarrow\phi$.
These regions of timelike identification are analogues of 
the `whiskers' of the closely related Nappi-Witten
cosmological spacetime~\cite{Nappi:1992kv}
discussed recently in~\cite{Elitzur:2002rt}
(indeed, the labelling of the slices 
in figure~\BTZslices\ is adopted from~\cite{Elitzur:2002rt}).  
Roughly, the gauging performed in the Nappi-Witten
geometry removes the BTZ radial direction from the 
Kruskal region of the global geometry, and the angular direction
from the `whisker' region.
In~\cite{Elitzur:2002rt} it was argued that 
observables are naturally defined in these regions,
in contrast to the conventional view
(\cf~\cite{Hawking:1992nk,Banados:1993gq})
that such regions are somehow forbidden or unstable.
These regions are certainly present if we regard
the BTZ geometry as an orbifold of global $AdS_3$.
In a sense, these regions are built-in by the identification
and the nature of the field configurations contributing
to the path integral; we see no obvious reason to
exclude them.  Again, making sense of them (or not)
in the 1+1 dimensional boundary CFT will go a long way
toward settling the question of their permissibility
in a quantum theory of gravity.

\vskip 2cm
\noindent
{{{\bf Acknowledgments}}}:
We thank
D. Kutasov,
J. Maldacena,
and 
S. Mathur
for discussions; EM wishes to thank LPTHE, Universit\'e
Pierre et Marie Curie, for its hospitality during the
completion of the manuscript.
This work was supported by DOE grant DE-FG02-90ER-40560.
 

\newpage
 

\providecommand{\href}[2]{#2}\begingroup\raggedright\endgroup

\end{document}